\documentclass[journal=jacsat,manuscript=article]{achemso}
\setkeys{acs}{articletitle = true}

\usepackage[version=3]{mhchem} % Formula subscripts using \ce{}
\usepackage[T1]{fontenc}       % Use modern font encodings
\usepackage{amssymb}           % smaller or approximately equivalent etc.
\author{Peilong Hou}
\affiliation[OS]{Institut f\"ur Chemie neuer Materialien, Universit\"at Osnabr\"uck, Barbarastr.7, 49076 Osnabr\"uck, Germany}
\author{Weijia Han}
\affiliation[OS]{Institut f\"ur Chemie neuer Materialien, Universit\"at Osnabr\"uck, Barbarastr.7, 49076 Osnabr\"uck, Germany}
\author{Michael Philippi}
\affiliation[OS]{Institut f\"ur Chemie neuer Materialien, Universit\"at Osnabr\"uck, Barbarastr.7, 49076 Osnabr\"uck, Germany}
\author{Helmut Sch\"afer}
\affiliation[OS]{Institut f\"ur Chemie neuer Materialien, Universit\"at Osnabr\"uck, Barbarastr.7, 49076 Osnabr\"uck, Germany}
\author{Martin Steinhart}
\email{martin.steinhart@uos.de}
\title[An \textsf{achemso} demo]
{Nanostructured submicron block copolymer dots by sacrificial stamping: a potential preconcentration platform for locally resolved sensing, chemistry and cellular interactions}

\begin{document}


%\begin{tocentry}
 %\begin{figure}[H]
%\centerline{\includegraphics{Figures/TOC.jpg}}
	%\caption{}
	%\label{TOC}
%\end{figure}
%\end{tocentry}


\begin{abstract}
Classical contact lithography involves patterning of surfaces by embossing or by transfer of ink. We report direct lithographic transfer of parts of sacrificial stamps onto counterpart surfaces. Using sacrificial stamps consisting of the block copolymer polystyrene-\textit{block}-poly(2-pyridine) (PS-\textit{b}-P2VP), we deposited arrays of nanostructured submicron PS-\textit{b}-P2VP dots with heights of $\sim$100 nm onto silicon wafers and glass slides. The sacrificial PS-\textit{b}-P2VP stamps were topographically patterned with truncated-pyramidal contact elements and penetrated by spongy-continuous nanopore systems. The spongy nature of the sacrificial PS-\textit{b}-P2VP stamps supported formation of adhesive contact to the counterpart surfaces and the rupture of the contact elements during stamp retraction. The submicron PS-\textit{b}-P2VP dots generated by sacrificial stamping can be further functionalized; examples include loading submicron PS-\textit{b}-P2VP dots with dyes and attachment of gold nanoparticles to their outer surfaces. The arrays of submicron PS-\textit{b}-P2VP dots can be integrated into setups for advanced optical microscopy, total internal reflection fluorescence microscopy or Raman microscopy. Arrays of nanostructured submicron block copolymer dots may represent a preconcentration platform for locally resolved sensing and locally resolved monitoring of cellular interactions or might be used as microreactor arrays in lab-on-chip configurations.          
\end{abstract}


\section{Introduction}
Chemical and topographic high-throughput patterning of surfaces by lithographic stamping is key to the preparation of a broad range of functional materials and components\cite{CL_Nie2008}. Nanoimprint lithography\cite{CL_Traub2016,CL_Guo2007} involves embossing of plastically deformable surfaces or surface coatings, often consisting of polymeric materials, with hard stamps. Classical soft lithography with elastomeric stamps \cite{CL_Xia1998a,CL_Qin2010} including approaches such as microcontact printing \cite{CL_Perl2009,CL_Kaufmann2010} and polymer pen lithography \cite{CL_Huo2008,CL_Braunschweig2009,CL_Carbonell2017} involves the transfer of molecules adsorbed on the stamp surface to a counterpart surface, on which consequently thin ink layers are deposited. Lithographic approaches that combine deposition of materials and topographic patterning may involve different types of capillary force lithography \cite{CL_Suh2009,CL_Ho2015}, insect-inspired capillary submicron stamping \cite{IICN_Han2018a}, wet lithography \cite{CL_Cavallini2009}, and electrochemical lithography\cite{CL_Simeone2009}. The stamping of functional materials characterized by complex molecular and/or mesoscopic architectures using classical soft lithography has remained demanding. Preformed nanoparticles having the desired functionality may be assembled on a first substrate, then transferred to a stamp and finally stamped onto a second substrate \cite{CL_Lee2007,CL_Cucinotta2009}. In the case of nanoparticles characterized by complex mesoscopic morphologies, such as mesoporous silica nanoparticles, deposition onto surfaces typically requires complex bonding chemistry \cite{NP_arrays_Kehr2010,NP_arrays_Wang2010,NP_arrays_Yoon2007,NP_arrays_Ruiz2006,NP_arrays_Lee2005}; spatial distribution and ordering of the mesoporous silica nanoparticles are difficult to control. 

Block copolymers (BCPs) are a particularly interesting class of materials because they may combine the specific properties of their chemically distinct blocks. In the case of the BCP polystyrene-\textit{block}-poly(2-vinylpyridine) (PS-\textit{b}-P2VP) the nonpolar PS blocks may serve as rigid glassy scaffold. The polar P2VP blocks can be functionalized taking advantage of the presence of a pyridyl group in each P2VP repeat unit. Also, reversible swelling of the P2VP domains can be controlled \textit{via} the pH value \cite{IICN_Wang2007}. BCPs themselves have been employed as structure-directing agents \cite{NP_arrays_Bang2009,NP_arrays_Lohmuller2011}. Also, either solid or mesoporous BCP nanoparticles are accessible by different solution-based preparative approaches\cite{NP_arrays_Deng2012,NP_arrays_Jin2014,NP_arrays_Fan2014,NP_arrays_Higuchi2017}. Lithographic deposition of BCPs by a parallel lithographic process would be a powerful approach to pattern and to functionalize surfaces but has yet not been established. 

\begin{figure}[htbp]
	\centering
	\includegraphics[width=0.5\textwidth]{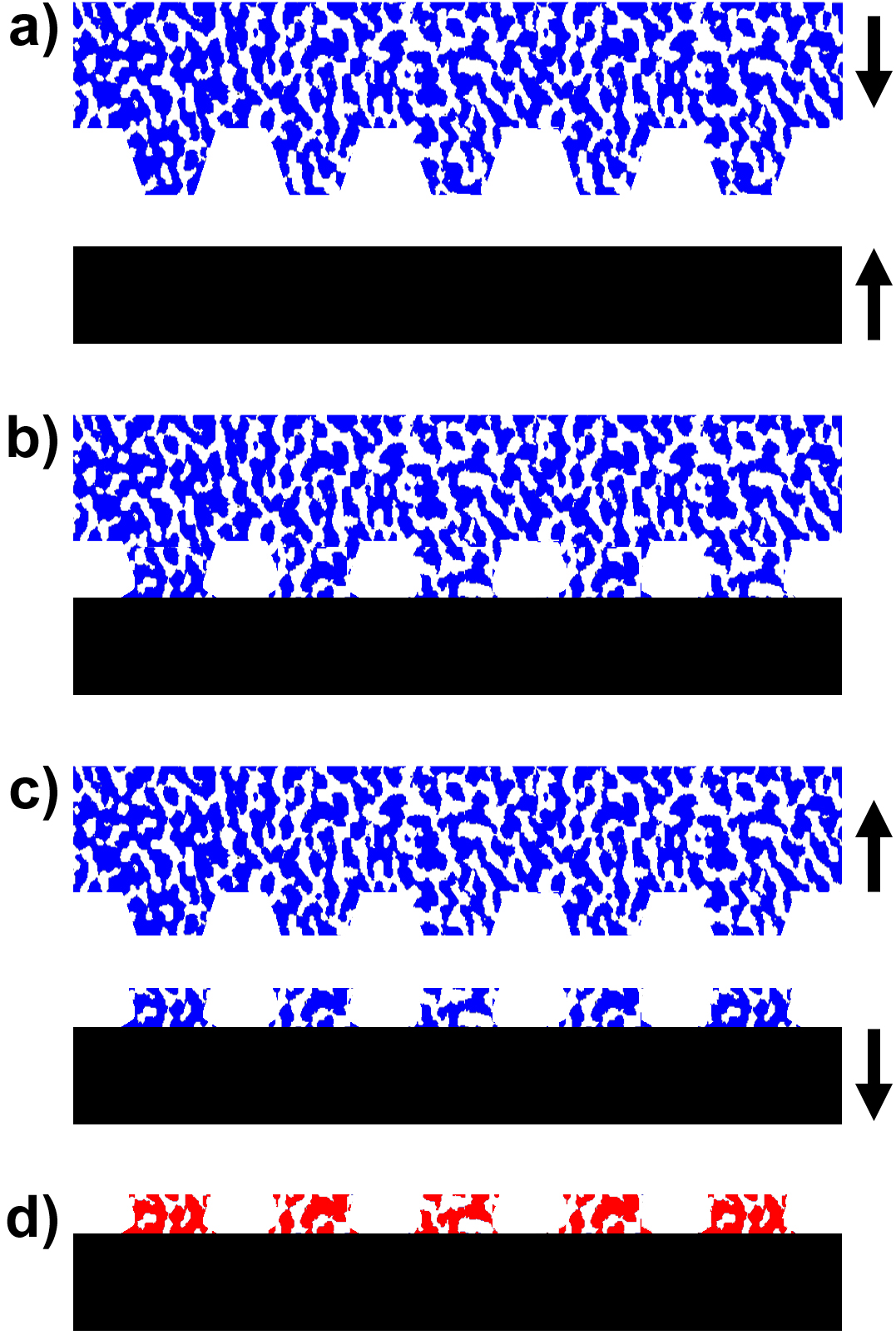}	
	\caption{Sacrificial stamping of arrays of submicron PS-\textit{b}-P2VP dots. a) A monolithic nanoporous PS-\textit{b}-P2VP stamp topographically patterned with truncated-pyramidal contact elements (blue) is pressed against a counterpart surface (black) located on a balance. b) The contact elements of the stamp form contact with the counterpart surface under a load controlled \textit{via} the mass displayed by the balance. c) Upon retraction of the nanoporous PS-\textit{b}-P2VP stamp, the contact elements rupture in such a way that their tips remain attached to the counterpart surface as submicron PS-\textit{b}-P2VP dots (blue). The PS-\textit{b}-P2VP, which the residual PS-\textit{b}-P2VP stamp consists of, can be recycled. d) The submicron PS-\textit{b}-P2VP dots thus deposited onto the counterpart surface can be further modified so that functionalized submicron PS-\textit{b}-P2VP dots (red) are obtained.}
\label{scheme_stamping}
\end{figure}

Here, we report sacrificial stamping to generate arrays of nanostructured submicron dots consisting of PS-\textit{b}-P2VP. Sacrificial stamping involves the lithographic transfer of material the sacrificial stamp itself consists of to a counterpart surface. A sacrificial PS-\textit{b}-P2VP stamp topographically patterned with contact elements is approached to a counterpart surface (Figure \ref{scheme_stamping}a). The PS-\textit{b}-P2VP stamp is then pressed against the counterpart surface so that tight adhesive contact between contact elements and counterpart surface forms (Figure \ref{scheme_stamping}b). Upon retraction of the sacrificial PS-\textit{b}-P2VP stamp, the contact elements rupture so that the layer of the contact elements in intimate contact with the counterpart surface remains attached to the latter (Figure \ref{scheme_stamping}c). As a result, parts of the sacrificial PS-\textit{b}-P2VP stamp are lithographically deposited onto the counterpart surface. The deposited submicron PS-\textit{b}-P2VP dots can be further functionalized (Figure \ref{scheme_stamping}d).

\section{Results and discussion}
\subsection{Preparation of sacrificial PS-\textit{b}-P2VP stamps} 
\begin{figure}[htbp]
	\centering
	\includegraphics[width=0.8\textwidth]{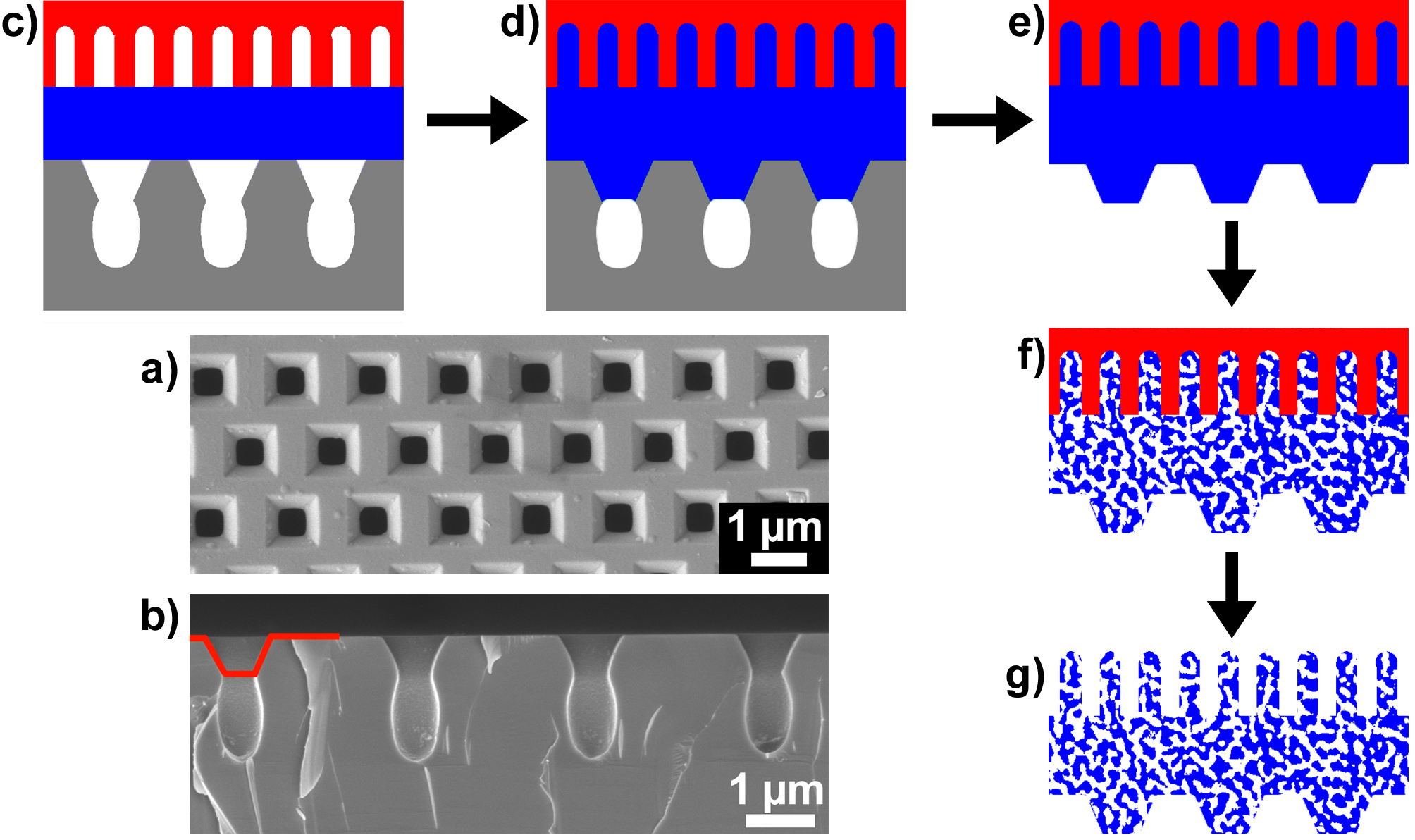}	
	\caption{Preparation of PS-\textit{b}-P2VP stamps. a) Top-view SEM image and b) cross-sectional SEM image of mSi in which the cross-section of a contact element of a sacrificial PS-\textit{b}-P2VP stamp is indicated by a red line. c)-g) Schematic diagram displaying the preparation of PS-\textit{b}-P2VP stamps. c) Molten PS-\textit{b}-P2VP (blue) is placed between self-ordered nanoporous AAO (red, on the top) and mSi (grey, at the bottom) modified with a perfluorinated silane. d) The PS-\textit{b}-P2VP melt partially infiltrates the macropores of the surface-modified mSi. e) After vitrification of the PS-\textit{b}-P2VP, the surface-modified mSi is non-destructively detached from the PS-\textit{b}-P2VP that is patterned with arrays of truncated pyramids at the initial positions of the mSi macropores. f) Continuous nanopore systems are generated in the PS-\textit{b}-P2VP by swelling-induced pore generation. The self-ordered AAO reinforces the PS-\textit{b}-P2VP specimens during swelling-induced pore generation and prevents bending and emergence of macroscopic waviness. The truncated pyramids obtained by molding the PS-\textit{b}-P2VP against surface-modified mSi are the contact elements of the obtained sacrificial PS-\textit{b}-P2VP stamps forming contact to the counterpart surfaces during sacrificial stamping. g) Optionally, the self-ordered AAO can be selectively etched.}
\label{stamp_prep.jpg}
\end{figure}

Sacrificial PS-\textit{b}-P2VP stamps were prepared as schematically displayed in Figure \ref{stamp_prep.jpg}. To topographically pattern the surfaces of the sacrificial PS-\textit{b}-P2VP stamps forming contact to the counterpart surfaces to be stamped, we molded molten PS-\textit{b}-P2VP against macroporous Si (mSi) \cite{SI_Lehmann1990,SI_Birner1998}. The mSi contained hexagonal arrays of macropores with a center-to-center distance of 1.5 $\mu$m (Figure \ref{stamp_prep.jpg}a). The inverse-pyramidal mouths of the mSi macropores (Figure \ref{stamp_prep.jpg}b) resulted from etch pits formed by wet-chemical pattern transfer following photolithographic prepatterning of silicon wafers. The positions of the etch pits defined the positions of the mSi macropores (pore depth $\sim$1.8 $\mu$m) generated by photoelectrochemical etching. The mSi macropores had a neck with a diameter of $\sim$530 nm directly below the inverse-pyramidal pore mouths. Below the neck, the mSi macropores widened and reached a diameter of ~710 nm (Figure \ref{stamp_prep.jpg}b). The surface of the mSi -- initially consisting of a thin native silica layer -- was coated with 1H,1H,2H,2H-perfluorodecyltrichlorosilane (PFDTS) following procedures reported elsewhere \cite{SI_Fadeev2000}. We melted PS-\textit{b}-P2VP sandwiched in between of surface-modified mSi and self-ordered anodic aluminum oxide (AAO) \cite{W_Masuda1998} (Figure \ref{stamp_prep.jpg}c). The self-ordered AAO containing arrays of nanopores with a pore diameter of 300 nm, a lattice period of 500 nm and a pore depth of 1.0 $\mu$m reinforced the PS-\textit{b}-P2VP specimens. In this way, bending of the PS-\textit{b}-P2VP specimens and formation of undulations during the preparation of the sacrificial PS-\textit{b}-P2VP stamps was prevented. During the annealing of the PS-\textit{b}-P2VP the low surface energy of the surface-modified mSi prevented complete infiltration of the mSi macropores; only the inverse-pyramidal pore mouths were filled with PS-\textit{b}-P2VP. Furthermore, the neck of the mSi macropores is an entropic barrier to infiltration; overcoming this barrier would require entropically unfavorable stretching of the PS-\textit{b}-P2VP chains (Figure \ref{stamp_prep.jpg}d). After cooling to room temperature, the surface--modified mSi was non--destructively detached from the vitrified PS-\textit{b}-P2VP specimens. As a result, PS-\textit{b}-P2VP films with arrays of truncated pyramids at the initial positions of the mSi macropores were obtained, while the mSi could be reused as mold (Figure \ref{stamp_prep.jpg}e). 

\begin{figure}[htbp]
	\centering
	
		a) \includegraphics[width=0.4\textwidth]{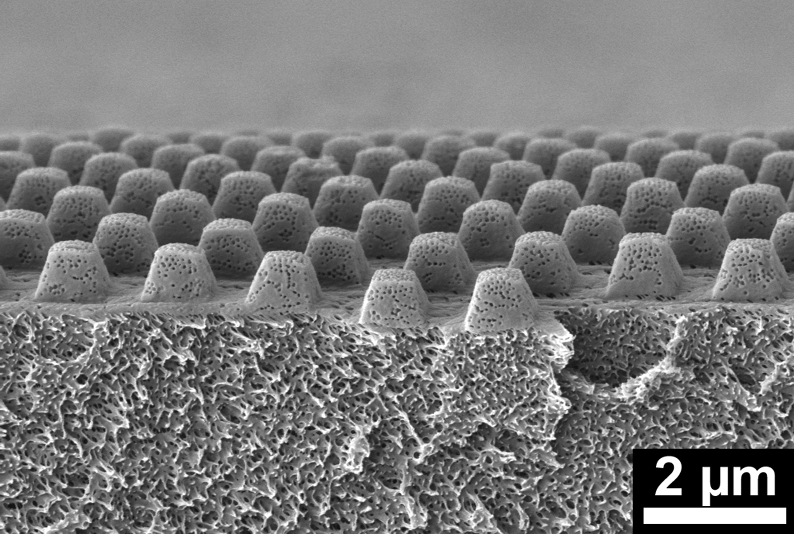}	
					
		b) \includegraphics[width=0.4\textwidth]{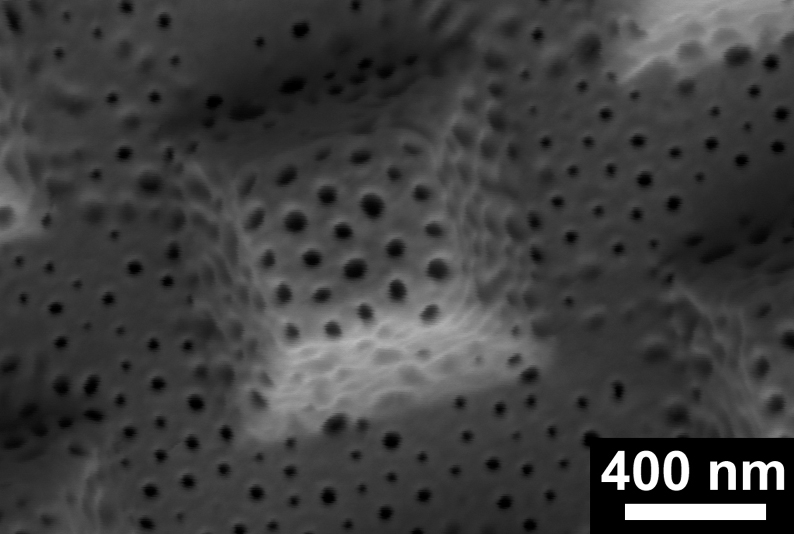}
		
		c) \includegraphics[width=0.4\textwidth]{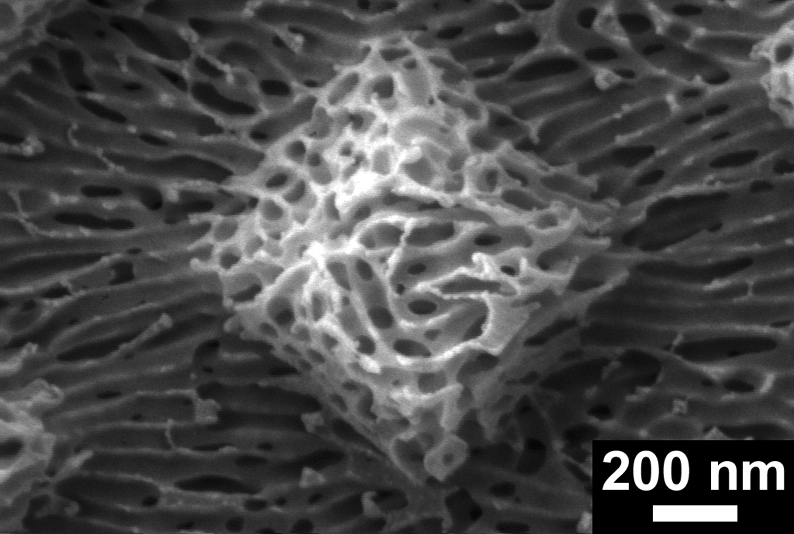}	
				
		d) \includegraphics[width=0.4\textwidth]{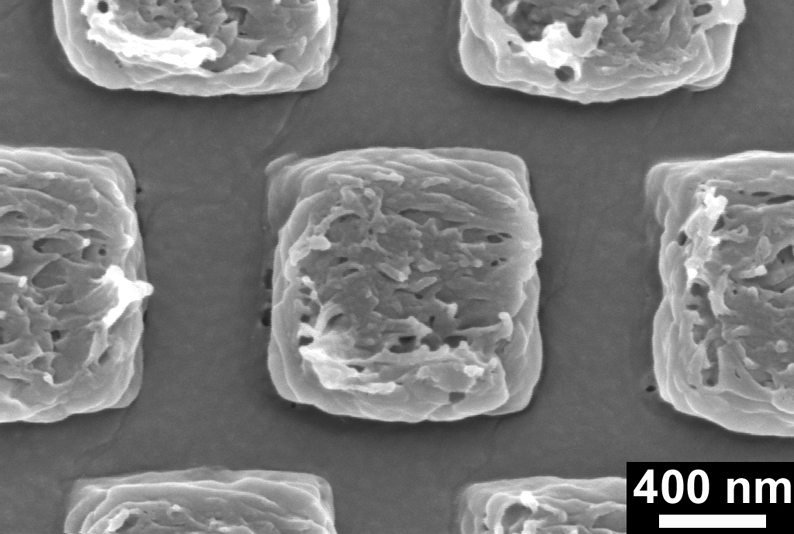}
	
	\caption{SEM images of PS-\textit{b}-P2VP stamps. a) Cross-section and b) single contact element prior to oxygen plasma treatment. c) Single contact element after oxygen plasma treatment (4 minutes at 100 W). d) Contact elements of a PS-\textit{b}-P2VP stamp after sacrificial stamping.}
\label{stamp_SEM}
\end{figure}

In the next step, we formed continuous nanopore systems in the topographically patterned PS-\textit{b}-P2VP specimens still attached to self-ordered AAO by swelling-induced pore generation with hot ethanol, which is a solvent selective to P2VP \cite{IICN_Wang2010,IICN_Wang2011,IICN_Eichler-Volf2016}. We applied a protocol established for the PS-\textit{b}-P2VP used here that results in the formation of continuous nanopore systems (Figure \ref{stamp_prep.jpg}f) characterized by a mean pore diameter of $\sim$40 nm, a specific surface area of 10 m$^2$/g, and a total pore volume of 0.05 cm$^3$/g \cite{IICN_Eichler-Volf2016}. Osmotic pressure drives the ethanol into the P2VP minority domains. The volumes of the P2VP minority domains increase because the P2VP blocks tend to maximize favorable interactions with ethanol molecules by assuming stretched conformations. The glassy PS matrix in turn undergoes structural reconstruction to accommodate the increased volumes of the P2VP minority domains swollen with ethanol. Bending and the development of macroscopic waviness related to volume expansion during swelling-induced pore generation in the PS-\textit{b}-P2VP specimens were prevented by the reinforcement with self-ordered AAO. Removal of the ethanol by evaporation results in entropic relaxation of the expanded P2VP blocks that transform to coils, while the glassy PS matrix fixates the reconstructed morphology. Consequently, nanopores with walls consisting of coiled P2VP blocks form in place of the expanded P2VP domains swollen with ethanol. To ensure that an isotropic pore network inside a sufficiently stable spongy-continuous scaffold forms, we used asymmetric PS-\textit{b}-P2VP containing PS as matrix component. The nanoporous PS-\textit{b}-P2VP specimens were then subjected to oxygen plasma. The sacrificial PS-\textit{b}-P2VP stamps obtained in this way, which are still connected to self-ordered AAO, may be used for sacrificial stamping and then recovered by again molding them against mSi and repeating steps c)--f) displayed in Figure \ref{stamp_prep.jpg}. Optionally, the self--ordered AAO can be selectively etched Figure (Figure \ref{stamp_prep.jpg}g) to facilitate the loading of the sacrificial stamps with additional components that may further modify and/or functionalize the stamped submicron PS-\textit{b}-P2VP dots. For the sacrificial stamping experiments reported here, we used freestanding and nanoporous sacrificial PS-\textit{b}-P2VP stamps with a thickness of $\sim$240 $\mu$m (Figure \ref{stamp_SEM}a-c). 

\subsection{Sacrificial stamping}
\begin{figure}[htbp]
	 %\captionsetup{labelformat=empty}
    \centering
		
		%a) \includegraphics[width=0.6\textwidth]{figs/particle_transfer_015_Kopie.jpg}	
					
		a) \includegraphics[width=0.6\textwidth]{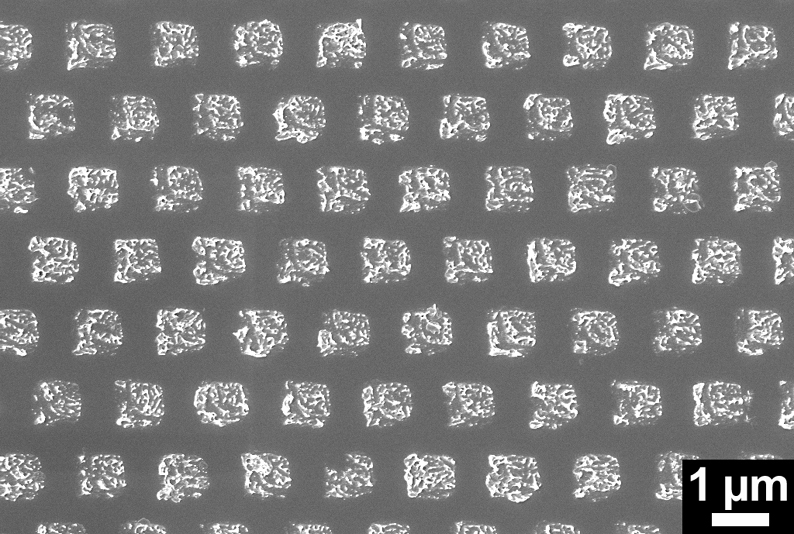}	
		
    b) \includegraphics[width=0.6\textwidth]{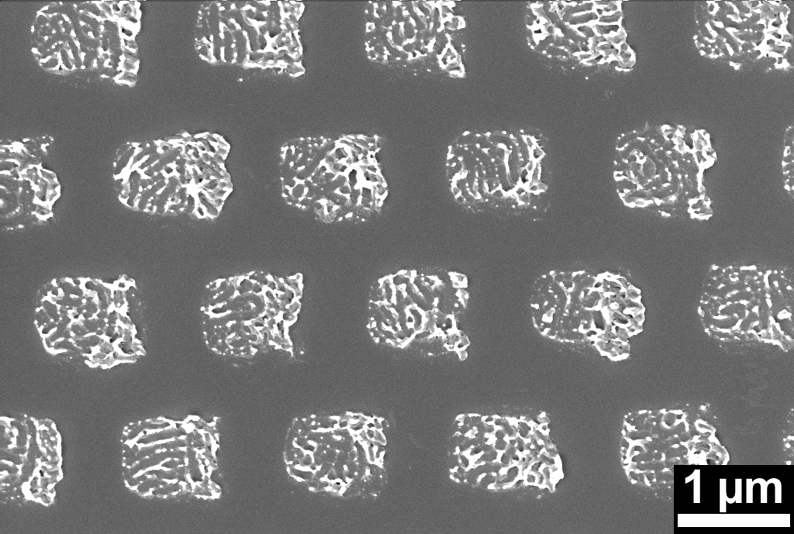}
    	
 %\caption{}
    	%\label{}
%\end{figure}
%\clearpage
%\begin{figure}
    %\captionsetup{labelformat=adja-page}
    %\ContinuedFloat
		c) \includegraphics[width=0.6\textwidth]{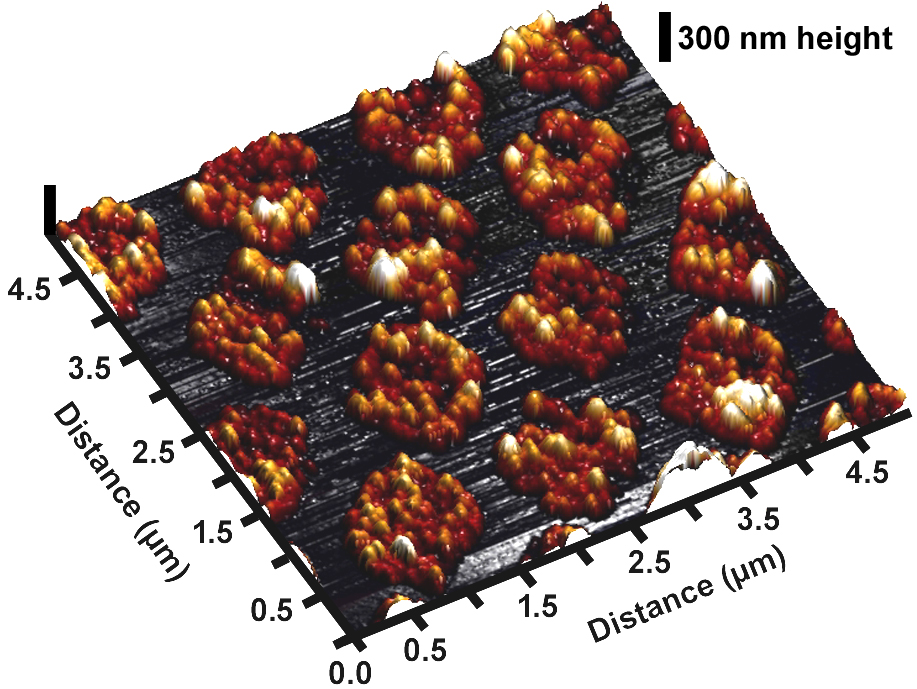}
\caption{Arrays of nanostructured submicron PS-\textit{b}-P2VP dots deposited on silicon wafers by sacrificial stamping. a), b) Scanning electron microscopy images; c) AFM topography image.}
\label{BCP_dots}
\end{figure}

We carried out sacrificial stamping by gluing sacrificial PS-\textit{b}-P2VP stamps onto steel cylinders that were pressed against counterpart surfaces located on a balance. The load was controlled \textit{via} the displayed mass. Using sacrificial PS-\textit{b}-P2VP stamps, we stamped hexagonal arrays of submicron PS-\textit{b}-P2VP dots with a lattice constant of 1.5 $\mu$m (Figure \ref{BCP_dots}; Supporting Figures S1 and S2) corresponding to the lattice constant of the mSi template initially used to topographically pattern the sacrificial PS-\textit{b}-P2VP stamps. Sacrificial stamping requires at first the formation of adhesive contact between the contact elements of the sacrificial PS-\textit{b}-P2VP stamps and the counterpart surfaces (Figure \ref{scheme_stamping}b). We enforced the formation of adhesive contact by controlled application of load. Moreover, formation of adhesive contact was promoted as follows.

i) The outer surfaces of the sacrificial PS-\textit{b}-P2VP stamps consisted of P2VP \cite{BCP-NP_Wang2008,BCP-NP_Wang2011}. The counterpart surfaces used here, silicon wafers covered by a native silica layer and glass slides, had hydroxyl-terminated surfaces. It was previously reported that strong attractive interactions between the pyridyl groups of P2VP and hydroxyl groups on the surface of silica substrates exist \cite{W_Roth2007}. Hence, specific chemical interactions between the P2VP surfaces of the contact elements of the sacrificial PS-\textit{b}-P2VP stamps and the hydroxyl-terminated counterpart surfaces enhance adhesion. 

ii) The sacrificial PS-\textit{b}-P2VP stamps were topographically patterned with hexagonal arrays of contact elements having the shape of truncated pyramids (cf. Figure \ref{stamp_SEM}a-c). The truncated pyramids with a height of $\sim$670 nm had flat, square-shaped upper surfaces with edge lengths of $\sim$550 nm (Figure \ref{stamp_SEM}b and c) that were congruent to the cross sectional areas of the macropore necks of the mSi (Figure \ref{stamp_prep.jpg}a). These flat contact surfaces of the contact elements facilitated formation of tight adhesive contact to counterpart surfaces as compared to, for example, contact elements with hemispherical tips. Thus, the quadratic shape of the contact surfaces of the contact elements was reproduced by the stamped PS-\textit{b}-P2VP dots

iii) While the counterpart surfaces can be considered as rigid and non-deformable, the spongy sacrificial PS-\textit{b}-P2VP stamps are deformable. As discussed above, the contour of the submicron PS-\textit{b}-P2VP dots was roughly rectangular, such as the contact surfaces of the contact elements of the sacrificial PS-\textit{b}-P2VP stamps. However, the edge lengths of the submicron PS-\textit{b}-P2VP dots amounted to $\sim$900 nm and exceeded, therefore, the edge lengths of the contact surfaces of the contact elements, as apparent from Figure \ref{stamp_SEM}b and c, by $\sim$65\%. Hence, in the course of sacrificial stamping the contact elements were compressed. Sacrificial stamping converted the open spongy morphology of the sacrificial PS-\textit{b}-P2VP stamps into a more densified morphology (Figure \ref{stamp_SEM}d). The areas of the densified contact elements matched those of the stamped submicron PS-\textit{b}-P2VP dots. The deformability of the contact elements increases the actual contact area between a contact element and the counterpart surface, which in turn results in enhanced adhesion per contact element. 

Secondly, the contact elements must rupture upon retraction of the sacrificial PS-\textit{b}-P2VP stamps. As a result, the parts of the contact elements in adhesive contact with the counterpart surfaces remain attached to the latter after removal of the sacrificial PS-\textit{b}-P2VP stamps (Figure \ref{scheme_stamping}c). The height of the stamped submicron PS-\textit{b}-P2VP dots ranged from $\sim$100 nm to $\sim$200 nm (Figure \ref{BCP_dots}c and Supporting Figure S3). The roughness of the surface topography of the submicron PS-\textit{b}-P2VP dots reflects the nanoporous-spongy nature of the sacrificial PS-\textit{b}-P2VP stamps. The crazing behavior of thin PS-\textit{b}-P2VP films has been reported to be complex \cite{BCP-NP_Lee2005,BCP-NP_Gurmessa2015}. Nevertheless, we assume that the following aspects promote the rupture of the contact elements. 

i) The spongy-nanoporous morphology of the sacrificial PS-\textit{b}-P2VP stamps reduces their tensile strength owing to confinement-induced reduction in the volume density of intermolecular entanglements. The viscosity of polymers as well as their behavior when subjected to stress is crucially related to the presence of intermolecular entanglements. The molecular weight of the PS blocks forming the majority component of the PS-\textit{b}-P2VP amounts to 101000 g/mol and lies, therefore, well above the threshold value $M_c$ of PS for the occurrence of intermolecular entanglements ($M_c$ $\approx$ 31200 g/mol at 490 K) \cite{W_Fetters1999}. However, PS chains in thin PS homopolymer films have been reported to be less entangled than in the bulk, corresponding to a confinement-induced increase in $M_c$ \cite{W_Brown1996,W_Si2005}. The nanopore walls in the sacrificial PS-\textit{b}-P2VP stamps have thicknesses in the 100-nm-range and below so that geometric restrictions similar to those in thin PS films apply. Moreover, the high incompatibility of the PS and P2VP blocks imposes additional constraints on the formation of intermolecular entanglements. Hence, the volume density of intermolecular entanglements within the PS domains in the PS-\textit{b}-P2VP stamps should be lower than in bulk PS-\textit{b}-P2VP. 

ii) Prior to sacrificial stamping, the sacrificial PS-\textit{b}-P2VP stamps were subjected to oxygen plasma treatment. The oxygen plasma treatment increased the areas of the nanopore openings at the surface of the sacrificial PS-\textit{b}-P2VP stamps, as obvious from a comparison of Figure \ref{stamp_SEM}b and Figure \ref{stamp_SEM}c. The oxygen plasma also cleaves some PS-\textit{b}-P2VP chains close to the surface of the sacrificial PS-\textit{b}-P2VP stamps into shorter segments. It is a straightforward assumption that cleavage of PS-\textit{b}-P2VP chains close the the surface of the contact elements and, therefore, close to the contact interface between contact elements and counterpart surface facilitates the rupture of the contact elements. 

\subsection{Modification of PS-\textit{b}-P2VP dots}
\begin{figure}[htbp]
    \centering
				a) \includegraphics[width=0.6\textwidth]{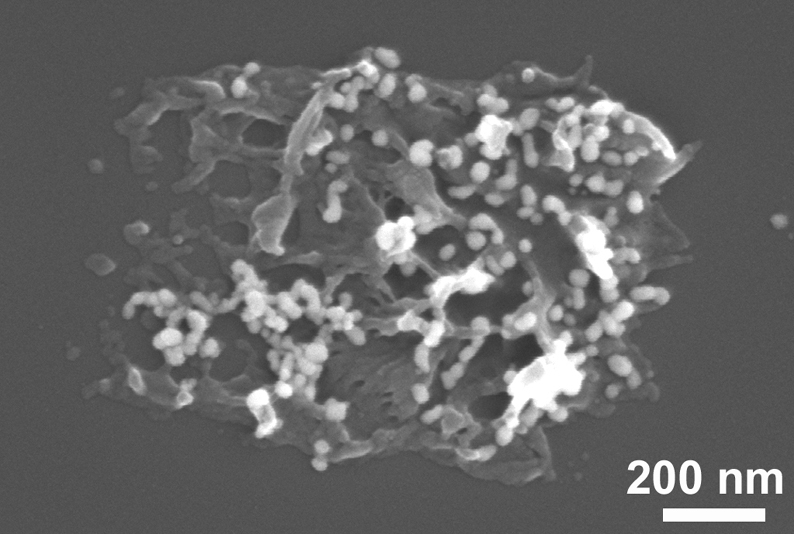}	
					
		b) \includegraphics[width=0.6\textwidth]{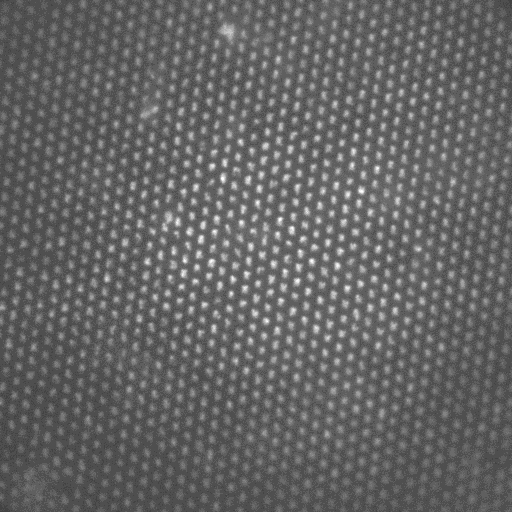}	
    	
 \caption{Functionalized submicron PS-\textit{b}-P2VP dots obtained by sacrificial stamping. a) Submicron PS-\textit{b}-P2VP dot on a Si wafer functionalized with gold nanoparticles. b) TIRFM image (edge length 54.5 $\mu$m) of the fluorescence of submicron PS-\textit{b}-P2VP dots containing rhodamine B stamped on a glass slide.}
\label{modifications}
\end{figure}

Since the outer surfaces of the submicron PS-\textit{b}-P2VP dots consist of P2VP blocks, the pyridyl groups of the latter are exposed to the environment and can be used for further functionalization. As example, we attached citrate--stabilized gold nanoparticles (AuNPs) with a diameter of 35 nm to the submicron PS-\textit{b}-P2VP dots by immersing arrays of submicron PS-\textit{b}-P2VP dots on Si wafers into AuNP suspensions with a pH value of 3.5. The negatively charged AuNPs adhered to partially protonated pyridyl moieties by van der Waals interactions and ionic interactions (P2VP is protonated at pH	$\leq$ 4.1 \cite{IICN_Wang2007}). Evaluation of 10 submicron PS-\textit{b}-P2VP dots revealed that on average 69 $\pm$ 13 AuNPs were bound to a submicron PS-\textit{b}-P2VP dot (Figure \ref{modifications}a). It is also possible to incubate the sacrificial PS-\textit{b}-P2VP stamp prior to sacrificial stamping with a functional material and to transfer the latter along with the submicron PS-\textit{b}-P2VP dots to a counterpart surface. As example, we filled a sacrificial PS-\textit{b}-P2VP stamp with a rhodamine B solution and let the solvent evaporate. Then, arrays of submicron PS-\textit{b}-P2VP dots containing rhodamine B were deposited on a glass slide by sacrificial stamping. The presence of rhodamine B was evidenced by total internal reflection fluorescence microscopy (TIRFM) imaging the fluorescence emission of rhodamine B (Figure \ref{modifications}b).     

\section{CONCLUSIONS}
In conclusion, we reported the ink-free lithographic transfer of parts of a sacrificial stamp to a counterpart surface. Using sacrificial stamps consisting of the block copolymer PS-\textit{b}-P2VP penetrated by spongy-continuous nanopore systems with truncated pyramids as contact elements, we stamped arrays of spongy submicron PS-\textit{b}-P2VP dots on silicon wafers and glass slides. Sacrificial stamping requires that adhesion of the stamp's contact elements to the counterpart surface is stronger than cohesion within the stamp. The enhanced deformability of the sacrificial PS-\textit{b}-P2VP stamps originating from their spongy-nanoporous morphology facilitated formation of adhesive contact between contact elements and counterpart surfaces. Suppression of intermolecular entanglements induced by the confinement of the thin nanopore walls inside the sacrificial PS-\textit{b}-P2VP stamps, as well as chain cleavage by oxygen plasma treatment, supported the rupture of the contact elements upon stamp retraction. Taking into account that the underlying polymer physics is generic, we assume that the methodology reported here is applicable to any polymeric stamp with spongy morphology. The PS-\textit{b}-P2VP, which the sacrificial PS-\textit{b}-P2VP stamps consist of, is recyclable. The pyridyl groups of the P2VP blocks forming the outer surface of the submicron PS-\textit{b}-P2VP dots facilitated further functionalization of the latter, for example, with gold nanoparticles or with dyes. Arrays of submicron PS-\textit{b}-P2VP dots stamped onto glass slides can be used as substrates for advanced optical microscopy, such as TIRF microscopy, as well as for Raman microscopy. Potential applications of block copolymer dot arrays generated by sacrificial stamping may include locally resolved preconcentration sensing and locally resolved monitoring of cellular interactions preconcentrated at the submicron block copolymer dots. Moreover, sacrificial stamping may yield lab-on-chip configurations consisting of arrays of microreactors for locally confined chemical reactions.     

\section{MATERIALS AND METHODS}
\textit{Materials.} Macroporous silicon (product number 620514-W23) was provided by SmartMembranes GmbH (Halle (Saale), Germany). Self-ordered AAO with a pore diameter of 300 nm, a lattice period of 500 nm and a pore depth of 1.0 $\mu$m was prepared by anodizing aluminum chips with a diameter of 4 cm (Goodfellow, purity >99.99 \%) following procedures reported elsewhere \cite{W_Masuda1998}. The self-ordered AAO layer was connected to a $\sim$1000 $\mu$m thick Al substrate. Asymmetric PS-\textit{b}-P2VP (M$_n$(PS) = 101000 g/mol; M$_n$(P2VP) = 29000 g/mol; M$_w$/M$_n$(PS-\textit{b}-P2VP) = 1.60, volume fraction of P2VP 21\%; bulk period $\sim$51 nm) was obtained from Polymer Source Inc., Canada. Tetrachloroauric(III)acid (HAuCl$_4$), trisodium citrate and rhodamine B were purchased from Sigma-Aldrich. 1H,1H,2H,2H-perfluorodecyltrichlorosilane (PFDTS, 97 \%, stabilized with copper) was supplied by ABCR GmbH, Germany. The AuNPs were synthesized following procedures reported elsewhere \cite{NP_arrays_Xie2009,IICN_Xue2017}. 1 mL of an aqueous 3.88 $\cdot$ 10$^{-2}$ M citrate solution was added to 50 mL of a boiling solution of 0.01 wt-\% HAuCl$_4$ in water. The aqueous mixture with a pH value of 3.5 was boiled for 20 min under vigorous stirring and then cooled to room temperature. A silicon wafer stamped with submicron PS-\textit{b}-P2VP dots was dipped into the thus-obtained aqueous AuNP suspension for 1 h, followed by three washing steps with deionized water.  

\textit{Preparation of sacrificial PS-\textit{b}-P2VP stamps.} To modify the surface of the mSi with PFDTS, the mSi was first treated with a boiling mixture containing 98\% H$_2$SO$_4$ and 30\% H$_2$O$_2$ at a volume ratio of 7:3 for 30 minutes, followed by rinsing with deionized water and drying in an argon flow. Then, the mSi was coated with PFDTS by vapor deposition for 2 h at 85$^\circ$C and for 3 h at 130$^\circ$C following procedures reported elsewhere \cite{SI_Fadeev2000}. About 240 $\mu$m thick PS-\textit{b}-P2VP films were prepared by dropping a solution of 100 mg PS-\textit{b}-P2VP per mL tetrahydrofuran (THF) onto a silicon wafer. After the complete evaporation of THF, the PS-\textit{b}-P2VP films were detached by exposure to ethanol for 24 h at room temperature and sandwiched between self-ordered AAO and surface-modified mSi in such a way that the macropore openings of the self-ordered AAO and the surface-modified mSi were in contact with the PS-\textit{b}-P2VP (Figure \ref{stamp_prep.jpg}c). The PS-\textit{b}-P2VP was infiltrated at 220$^\circ$C for 4 h under vacuum while a load of ~160 mbar was applied. The PS-\textit{b}-P2VP was cooled to room temperature at a rate of --1 K/min and immersed into ethanol for $\sim$30 min. Swelling-induced pore generation was carried out in ethanol at 60$^\circ$C for 4h. The Al substrate connected to the self-ordered AAO layer was selectively etched with a solution of 100 mL 37\% HCl and 3.4 g CuCl$_2$ * 2 H$_2$O in 100 mL deionized water at 0$^\circ$C. Finally, the self-ordered AAO was removed by etching with 3 M NaOH$_{(aq)}$ followed by washing with deionized water. The sacrificial PS-\textit{b}-P2VP stamps were finally subjected to oxygen plasma at 100 W for 4 minutes using a plasma cleaner Femto (Diener electronic, Ebhausen, Germany).

\textit{Sacrificial stamping.} Silicon wafers were cut into small pieces with areas of 1 cm $\cdot$ 1 cm. Thus-obtained Si wafer pieces and glass slides used as substrates for sacrificial stamping were washed with acetone, dried in an argon flow and treated with oxygen plasma (100 W; 10 min) using a plasma cleaner Femto (Diener electronic, Ebhausen, Germany). The sacrificial PS-\textit{b}-P2VP stamp used to produce the sample shown in Figure \ref{modifications}b was immersed into a solution of 50 mmol/L rhodamine B in ethanol for 1h followed by three washing steps with deionized water and drying at 35$^\circ$ for 12h. Prior to sacrificial stamping, the PS-\textit{b}-P2VP stamps were attached to a home-made stamp holder made of stainless steel (length: 55 mm, diameter: 12 mm, mass: 40 g) using double sided adhesive tape. Sacrificial stamping was carried out manually while applying a pressure of $\sim$50 bar. The pressure was adjusted by carrying out sacrificial stamping on a balance and calculated from the displayed mass. The contact time amounted to $\sim$1 s.  

\textit{Characterization.} SEM investigations were carried out on a Zeiss Auriga microscope operated at an accelerating voltage of 3 kV. For SEM, the samples were sputter-coated with a 5 nm thick iridium layer. AFM measurements were conducted in semicontact mode using a NT-MDT NTEGRA device. The cantilevers had a nominal length of 95 $\mu$m, force constants of 3.1--37.6 N m$^{-1}$, and a resonance frequency of 256 kHz (within the range of 140--390 kHz). The tip radius was 10 nm. The AFM images were processed by using the software Nova Px. TIRFM was performed using an inverted microscope (IX71, Olympus) equipped with a motorized 4-line TIRF condenser (cellTIRF 4-Line system, Olympus), a 150fold oil immersion TIRF objective (UAPON 150xTIRF, NA 1.45 Olympus) and a 561 nm diode-pumped solid state laser (max. power 200 mW; Cobolt Jive 561, Cobolt, Sweden). Images were acquired by an electron multiplying back-illuminated frame transfer CCD camera (iXon Ultra 897, Andor). A fluorescence filter cube containing a polychroic beamsplitter (R405/488/561/647, Semrock) and a quad-band emission/blocking filter (FF01 446/523/600/677, Semrock) was used. For each sample, 500 frames were recorded with an exposure time of 31 ms, a cycle time of 67 ms and a laser power of 5 mW (power density approx. 150 W/cm$^2$).

\section{ASSOCIATED CONTENT}
\begin{suppinfo}
Figure S1: Large-area SEM image of an array of submicron PS-\textit{b}-P2VP dots on a Si wafer generated by sacrificial stamping. Figure S2: SEM image at intermediate magnification of an array of submicron PS-\textit{b}-P2VP dots on a Si wafer generated by sacrificial stamping. Figure S3: AFM topography line profile of submicron PS-\textit{b}-P2VP dots deposited on a Si wafer by sacrificial stamping.
\end{suppinfo}

\section{AUTHOR INFORMATION}
\subsection{Corresponding Author}
*E-mail: martin.steinhart@uos.de (M. S.).

\subsection{ORCID}
Martin Steinhart: 0000-0002-5241-8498

\subsection{Notes}
The authors declare no competing financial interest.

\begin{acknowledgement}
The authors thank the European Research Council (ERC-CoG-2014, project 646742 INCANA) for funding. The preparation of self-ordered AAO by C. Hess, I. Hanemann and C. Schulz-K\"olbel is gratefully acknowledged. 
\end{acknowledgement}

\bibliography{bib/BCPs_in_nanopores,bib/NP_arrays,bib/wetting,bib/Si,bib/contact_litho,bib/IICN,bib}

\newpage
\center
\includegraphics{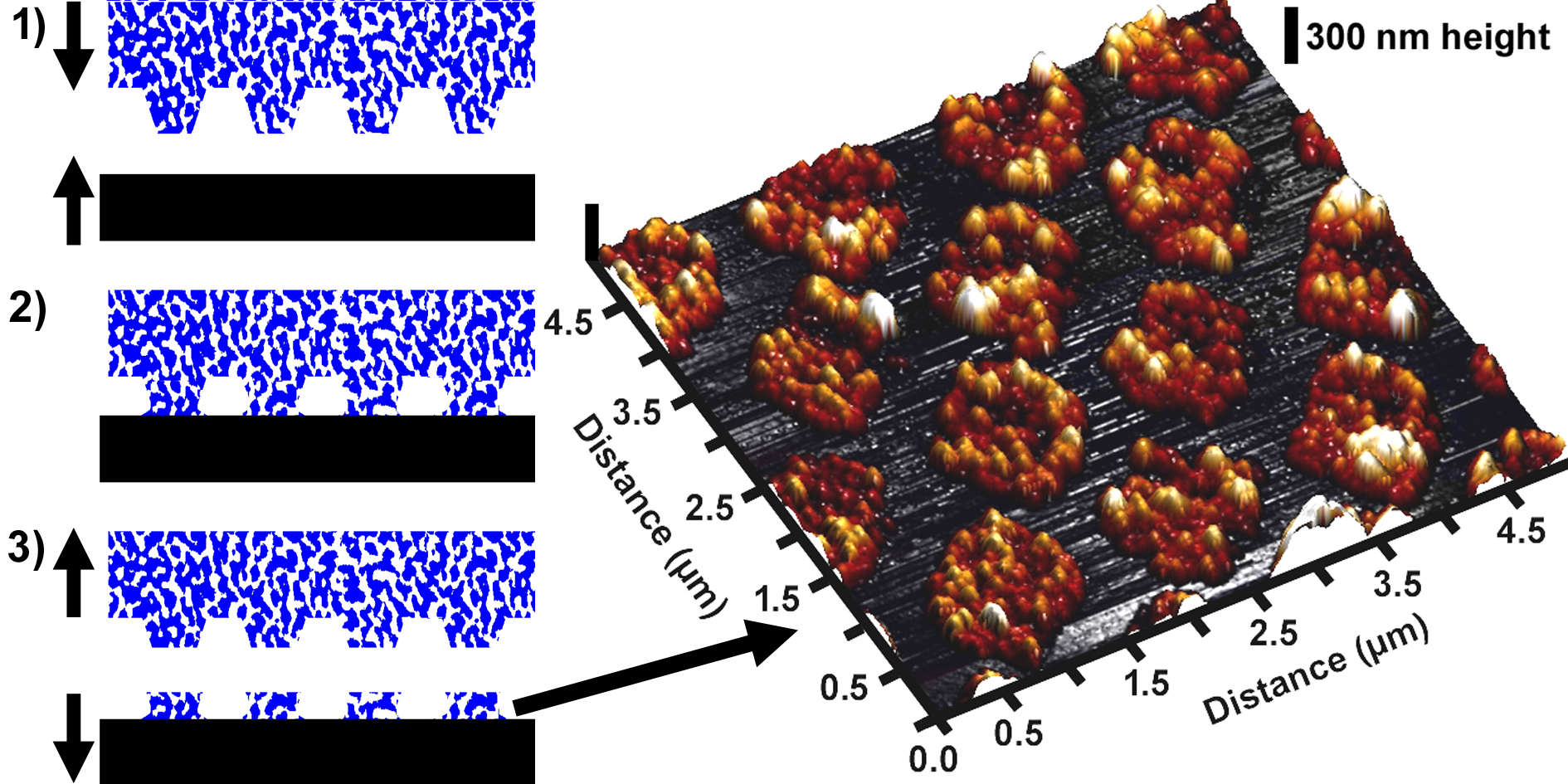}\\
\vspace{0.5cm}
Table of Contents graphics

\end{document}